\documentclass[journal=jpcbfk,manuscript=article]{achemso}

\usepackage{amsmath}
\usepackage{amssymb}
\usepackage{booktabs}
\usepackage{graphicx}
\usepackage{siunitx}
\usepackage{xcolor}
\usepackage{hyperref}
\usepackage{caption}
\DeclareSIUnit\angstrom{\text{\AA}}

\title{Calculations of the Krypton Phase Diagram and Novel Plasticity}

\author{Marcin Kirsz}
\email{marcin.kirsz@ed.ac.uk}

\author{Asuka Iwasaki}

\author{Graeme J. Ackland}
\email{gjackland@ed.ac.uk}
\phone{01316505299}

\affiliation{Centre for Science at Extreme Conditions, School of Physics and Astronomy, The University of Edinburgh, Edinburgh EH9 3FD, United Kingdom}

\SectionNumbersOn

\date{\today}

\begin{document}
\maketitle

\begin{abstract}

The phase diagram for Kr, as represented by the Tadah! two-body potential is shown to have face-centred cubic (fcc), hexahonal close packed (hcp), and body centred cubic (bcc) regions. It has been assembled by combining several methods: direct liquid--solid coexistence for the melt lines, Gibbs--Helmholtz integration and Clapeyron slopes for the bcc--fcc line, slab coexistence for the liquid--gas line, static zero-temperature relaxations for the crystals, and the quasiharmonic approximation for the low-temperature fcc--hcp windows. The bcc phase contains highly mobile ``greedy snake" defects, which suggests a reinterpretation of the melt-curve data: the anomaly observed may be due to the speckle method detecting the bcc-fcc boundary, not the melt curve.
While pair potentials have limitations, comparison with a foundation MACE model shows that a more flexible machine-learned model does not necessarily improve matters if inappropriately trained.
\end{abstract}

\section{Introduction}

Rare-gas solids are often regarded as the simplest materials for atomistic modelling. Unlike metals or covalent materials, their electrons can be easily assigned to atoms; unlike ionic solids, their forces are short-ranged; and unlike molecular systems, there is no separation of energy scales between intermolecular and intramolecular forces. Traditionally, simple two-body potentials such as Lennard--Jones have been assumed sufficient.

Experimentally, krypton is reported to have a simple phase diagram with an fcc crystal structure and liquid and gas phases.
However, there are some curiosities.
X-ray crystallographic studies have reinterpreted extensive stacking faults, which were previously identified as crystal defects\cite{van1996cross,sonnenblick1982growth}, as a high-pressure hcp phase, while Raman spectroscopy has also been used to study this proposed transition\cite{shimizu2009high}. Subtle anomalies in the EXAFS signal on pressurisation have also been reported\cite{rosa2018effect}. Melting studies\cite{Boehler2001melting} based on the speckle method report an anomaly in the melt curve at about \SI{50}{GPa}, where it departs from the typical Simon--Glatzel extrapolation.  The speckle method\cite{jeanloz1996melting} shines a coherent laser through the diamond anvil cell onto a solid surface to generate a grainy interference pattern from surface inhomogeneity. Heating causes a sudden loss of pattern correlation as the rigid structure collapses into a fluid liquid.  

The advent of machine-learning interatomic potentials has been accompanied by human unlearning of some of the principles of effective models for condensed matter. In particular, easily calculated quantities such as forces and energies compared with DFT have replaced measurable properties in the assessment of potential accuracy.
A low fitting error to energies and forces demonstrates only that the functional form is flexible enough to represent the training data. It does not demonstrate that the model is useful.
Unfortunately, this diagnostic does not provide evidence for more important quantities, such as phase stability, even though DFT is extremely reliable for this purpose.
There are few studies of full phase diagrams, but those that exist reveal unexpected and incorrect behaviour even for well-established potentials\cite{partay2021nested,loach2017stacking}.
Here we investigate the phase diagrams of machine-learned potentials for krypton.

The thermodynamic definition of a phase boundary is the line along which the Gibbs free-energy difference between the phases is zero. Statistical mechanics allows us to calculate the Gibbs free energy by summing over microstates, although this introduces the problem of defining which microstate is associated with which phase.

In some cases, one can take a representative sample of the entire partition function and look for discontinuities in its derivatives. This requires a sampling method which can evenly sample all the microstates.
Alternatively, one can devise a sampler which is confined to a single phase. Typically, the thermodynamic properties will converge much faster. Molecular dynamics (MD) and thermodynamic integration methods come into this category. For any given run, some diagnostic is needed to check that the intended phase has been preserved during sampling. This single-phase approach works well for equations of state, but to find phase transformations it must either be augmented by a calculation which gives the absolute free energy of each phase\cite{frenkel1984new,bruce1997lattice,bruce2000lattice,MorrisSong2002}, or model two phases simultaneously to converge
the free energy difference.

Here we compare two potentials for krypton: a two-body potential built using Tadah! and a many-body model built using MACE. Both models use a machine-learning framework to fit to {\it ab initio} accuracy: MACE uses standard DFT, while Tadah!Kr3b uses coupled-cluster data. This exemplifies a general trade-off: DFT can sample many-atom configurations but describes dispersion poorly, while coupled-cluster theory captures dispersion accurately but is affordable only for dimers and trimers.
Kr is believed to remain non-metallic up to \SI{316}{GPa}\cite{hama1989equation}.

The Tadah!Kr3b potential has been published previously\cite{iwasaki2025accurate,kirsz2025tadah} and is fitted directly to coupled-cluster calculations\cite{Jager2016}, as detailed in the Methods section. We choose the ``foundation'' MACE-MP-0 model\cite{Batatia2022,batatia2025foundation} (2023-12-03-mace-128-L1\_epoch-199.model), which ``can be applied out of the box as a starting or `foundation' model for any atomistic system of interest''\cite{batatia2025foundation}. 
Both Tadah!Kr3b and MACE are well interfaced with LAMMPS\cite{thompson2022lammps}, the molecular dynamics code we use here.

The Lennard--Jones potential is widely used for different rare-gas elements, although classically the models are identical apart from their length and energy units. Its phase diagram, with hcp stable at low temperature and fcc favoured on heating and compression, has already been widely studied\cite{jackson2002lattice,travesset2014phase,schultz2018comprehensive,schwerdtfeger2024hundred,wiebe2020phase}, and we do not repeat those calculations here. For krypton specifically, neutron-scattering data is better described by a longer-ranged and softer-cored model than the 12--6 form permits\cite{pruteanu2022krypton}.

There have been simulation reports of a bcc phase in Xe\cite{belonoshko2002molecular,belonoshko2006xenon}; we demonstrate here that it is also present in Kr.
We find an unexpected phase transition at low temperature with the Tadah!Kr3b potential. In a companion paper in this issue\cite{kirsz2026pitfalls}, we investigate the general question this raises: whether \emph{every} two-body potential must exhibit a solid--solid phase transition.

\section{Methods}

The key tool used here is molecular dynamics, which we use to sample enthalpy and density in various phases and for direct two-phase coexistence calculations of the melting and boiling curves. We also use static relaxation and the quasiharmonic approximation at low temperatures, and Frenkel--Ladd thermodynamic integration as an independent check on the relative stability of the solid phases.

Here we evaluate the phase diagram for krypton using the Tadah!Kr3b potential downloaded from the Tadah!Kr3b website. The calculations include:

Code versions used were Tadah! 1.3.0-beta.1, 
\subsection{Molecular dynamics}

We ran molecular dynamics using the LAMMPS package\cite{thompson2022lammps}, with a \SI{1}{fs} timestep and the tabulated Tadah!Kr3b potential. Static relaxations used conjugate-gradient minimisation, for which spline interpolation proved necessary.
Nose--Hoover thermostats (\SI{200}{fs} damping) and barostats (\SI{5}{ps} damping) were used. Typical runs used a \SI{50}{ps} NVT warm-up, \SI{500}{ps} of NPT equilibration, and \SI{2}{ns} production runs with around 1,000 atoms for the Clapeyron-slope simulations, extending to 500,000 atoms for the phase-coexistence and bcc-snake calculations. Pressure was averaged over three directions, except for the gas--liquid slab, for which only $P_{zz}$ was used to avoid surface-tension anomalies\cite{ackland1986semi}.

\subsection{Tadah!Kr3b Potential}

The training data set, to which the Tadah!Kr3b potential was fitted, consists of coupled-cluster interaction energies from J{\"a}ger et al. \cite{Jager2016}, computed at the CCSD(T) level in the complete-basis-set limit with corrections up to full quadruple excitations, core--core and core--valence correlation, and relativistic effects. They provide 36 data points for the potential energy between pairs of Kr atoms as a function of separation distance, $V_{2}(r)$, from $r=2.2$ to $15$~\AA. This source also provides 11 data points for $\Delta V_{3}(r)$, the nonadditive three-body energies of equilateral configurations of three Kr atoms, which are added as corrections to the two-body potential\cite{iwasaki2025accurate}.

\begin{equation}
    V'(r)= V_{2}(r) + \tfrac{1}{3}\Delta V_{3}(r).
\end{equation}

\begin{figure}[htbp]
\centering
\includegraphics[width=0.90\textwidth]{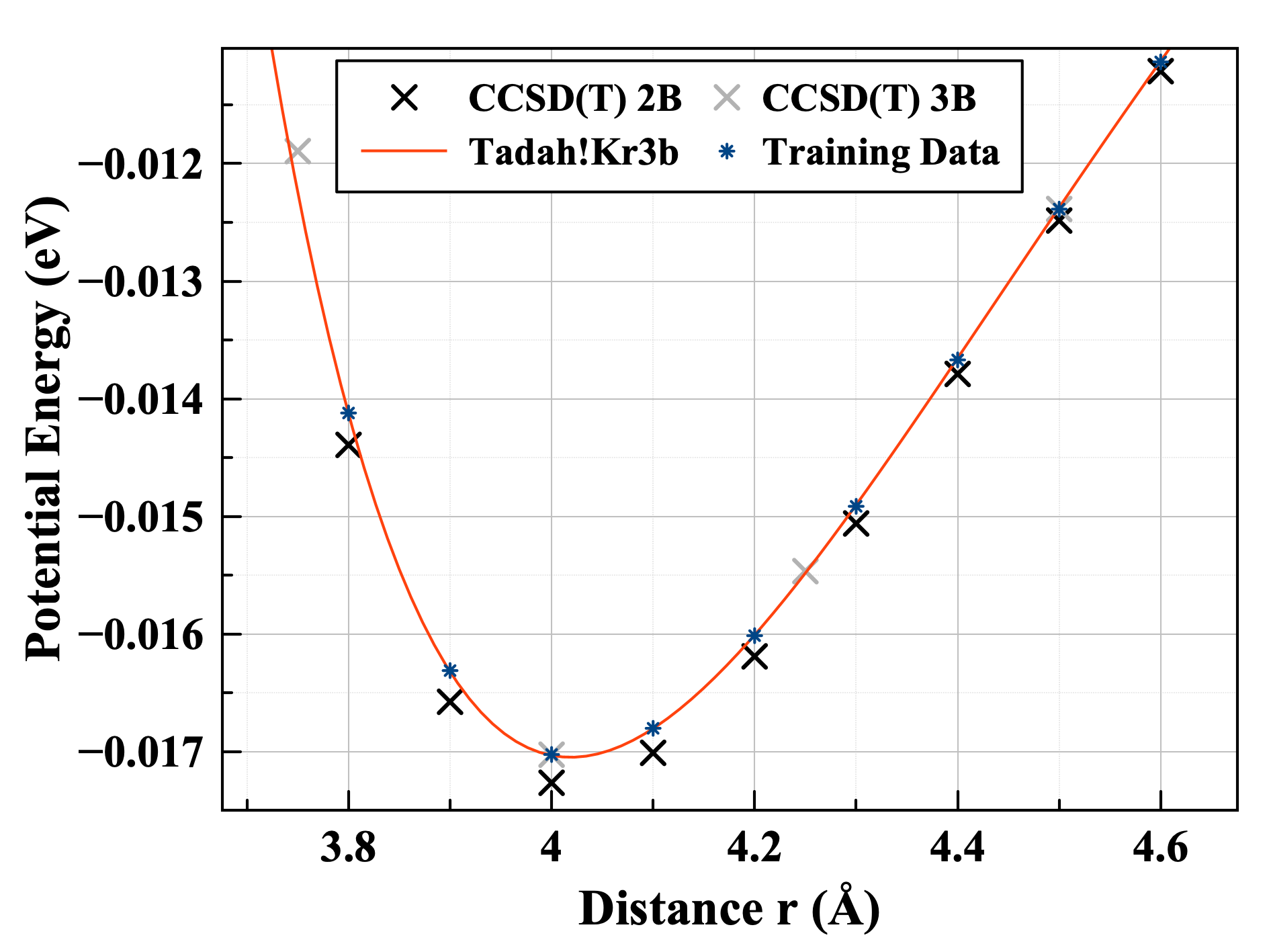}
\caption{Dimer potential-well region. The final training data set combines the original CCSD(T) two-body energies and three-body corrections. Interpolated Tadah!Kr3b 3B corrections were used for unmatched distances. The trained Tadah!Kr3b potential is also shown.}
\label{fig:training}
\end{figure}

Here $V'$ is the effective two-body interaction: the total energy of the equilateral trimer, $3V_2+\Delta V_3$, is recovered as a sum over its three bonds, and $V'$ is equivalently the energy per atom of the trimer.
These three-body energy corrections are provided over the range 2.5--6~\AA. The 3B energy corrections were fitted separately with Tadah! so that interpolated values could correct the 2B data at distances with no corresponding 3B value in the original CCSD(T) data. Figure~\ref{fig:training} shows the original CCSD(T) data points, the combined training data set, and the resulting Tadah!Kr3b potential around the well region.

Descriptors are the functions used to fit the training data. The Tadah!Kr3b potential comprises one ZBL screened-Coulomb function\cite{ziegler1985stopping}, $V_{ZBL}(r)$, in the repulsive short-range region and seven Gaussian-type functions in the well and attractive regions, giving the model energy
\begin{equation}
V(r) = w_0\, V_{\rm ZBL}(r)\, f_c^{(1)}(r) + \sum_{k=1}^{7} w_k\, e^{-\eta_k (r-r_k)^2}\, f_c^{(2)}(r),
\label{eq:Tadah!Kr3b_model}
\end{equation}
where the weights $w_k$ are linear regression coefficients, the centres $r_k$ and widths $\eta_k$ are machine-optimised hyperparameters, and $f_c^{(1)}$ and $f_c^{(2)}$ are cutoff functions taking the ZBL term smoothly to zero at short range and the Gaussian terms smoothly to zero at 12~\AA. The training data were augmented with pure ZBL energies for $0.5\le r\le2$~\AA. The rationale is that we trust the coupled-cluster data in the well and attractive regions, and we trust the ZBL form for core--core repulsion, but less so its universal screening function at intermediate separations: the fitted amplitude $w_0$ and the smooth join between the highest-energy coupled-cluster point (2.2~\AA, $\approx4.6$ eV) and the ZBL data allow the training data to decide. A previously-published Tadah!Kr3b fit\cite{iwasaki2025accurate} was left unconstrained below the shortest data point. The ZBL extension removes the resulting unphysical short-range behaviour, which matters for simulations at high pressure and temperature.

The Tadah!Kr3b potential was trained using Tadah!'s Hyperparameter Optimisation (HPO) algorithm. HPO is a feature of Tadah! which automatically optimises the descriptors' positions and widths (hyperparameters), as well as their weights (learned parameters), via a nested fitting procedure\cite{kirsz2025tadah}. Starting from a proposed set of hyperparameters, the inner loop trains the potential using a chosen regression method, in this case Bayesian linear regression, which assigns weights to the descriptors. The outer loop then evaluates the fitted model and adjusts the hyperparameters before calling the inner loop again. Over 100,000 optimisation cycles, HPO arrived at the Tadah!Kr3b potential about halfway through, after about five minutes.

\subsection{Clapeyron Slope Field}

Our sole technical innovation is the definition and use of the
``Clapeyron slope field'' $\mathcal{C}(T,P)$, which
generalises the Clausius--Clapeyron expression to all $(P,T)$ conditions for any pair of states:

\begin{equation}
\mathcal{C}(T,P) \equiv 
\left(\frac{dT}{dP}\right)_{\mathrm{\alpha\beta}} = \frac{T\,\Delta v(T,P)}{\Delta h(T,P)}= T\frac{v_\alpha(T,P)-v_\beta(T,P)}{h_\alpha(T,P)-h_\beta(T,P)}
\label{eq:slope_field_definition}
\end{equation}
where $\alpha$ and $\beta$ label the two states, and $v$ and $h$ are the volume and enthalpy per atom.
This is easily evaluated using paired, independent, single-phase NPT MD calculations at any chosen set of $(P,T)$.

Along the line $\Delta G =0$ this reduces to the familiar Clausius--Clapeyron equation for the slope of the phase boundary.
Unfortunately, a simple enthalpy calculation does not determine
$\Delta G_{\alpha\beta}$, but Eq.~\ref{eq:slope_field_definition} can be generalised across the whole phase diagram.
It enables Gibbs--Duhem integration\cite{KofkeMP1993,KofkeJCP1993} along a phase boundary, but is easier to calculate from molecular dynamics than an equipotential line.
To see what this means, consider two candidate phases, $\alpha$ and $\beta$, with Gibbs free-energy difference
\begin{equation}
\Delta g(T,P) \equiv g_\beta(T,P) - g_\alpha(T,P).
\end{equation}
$\Delta g$ is zero on the coexistence curve. 
The exact differential of $g$ is
$
dg = -s\,dT + v\,dP,
$
where $s$ and $v$ are entropy and volume per particle. Therefore
\begin{equation}
d(\Delta g) = -\Delta s\,dT + \Delta v\,dP,
\end{equation}
with
$
\Delta s \equiv s_\beta - s_\alpha;
\,
\Delta v \equiv v_\beta - v_\alpha.
$
Equivalently, in partial-derivative form,
\begin{equation}
\left(\frac{\partial \Delta g}{\partial T}\right)_P = -\Delta s,
\qquad
\left(\frac{\partial \Delta g}{\partial P}\right)_T = \Delta v.
\end{equation}

On the phase line only ($\Delta g=0$), this gives the usual Clausius--Clapeyron slope of the coexistence curve:
\begin{equation}
\left(\frac{dT}{dP}\right)_{\mathrm{coex}} = \frac{\Delta v}{\Delta s} = \frac{T\Delta v}{\Delta h} .
\label{eq:clapeyron_entropy_form}
\end{equation}
In general, using $h = g + Ts$, so that
$\Delta h = \Delta g + T\Delta s$,
 the slope field definition can be rewritten as:
\begin{equation}
\mathcal{C}(T,P) \equiv \frac{\Delta v(T,P)}{\Delta s(T,P) + \Delta g(T,P)/T}
\label{eq:slope_field_definition2}
\end{equation}

\begin{figure}[htbp]
\centering
\includegraphics[width=0.98\textwidth]{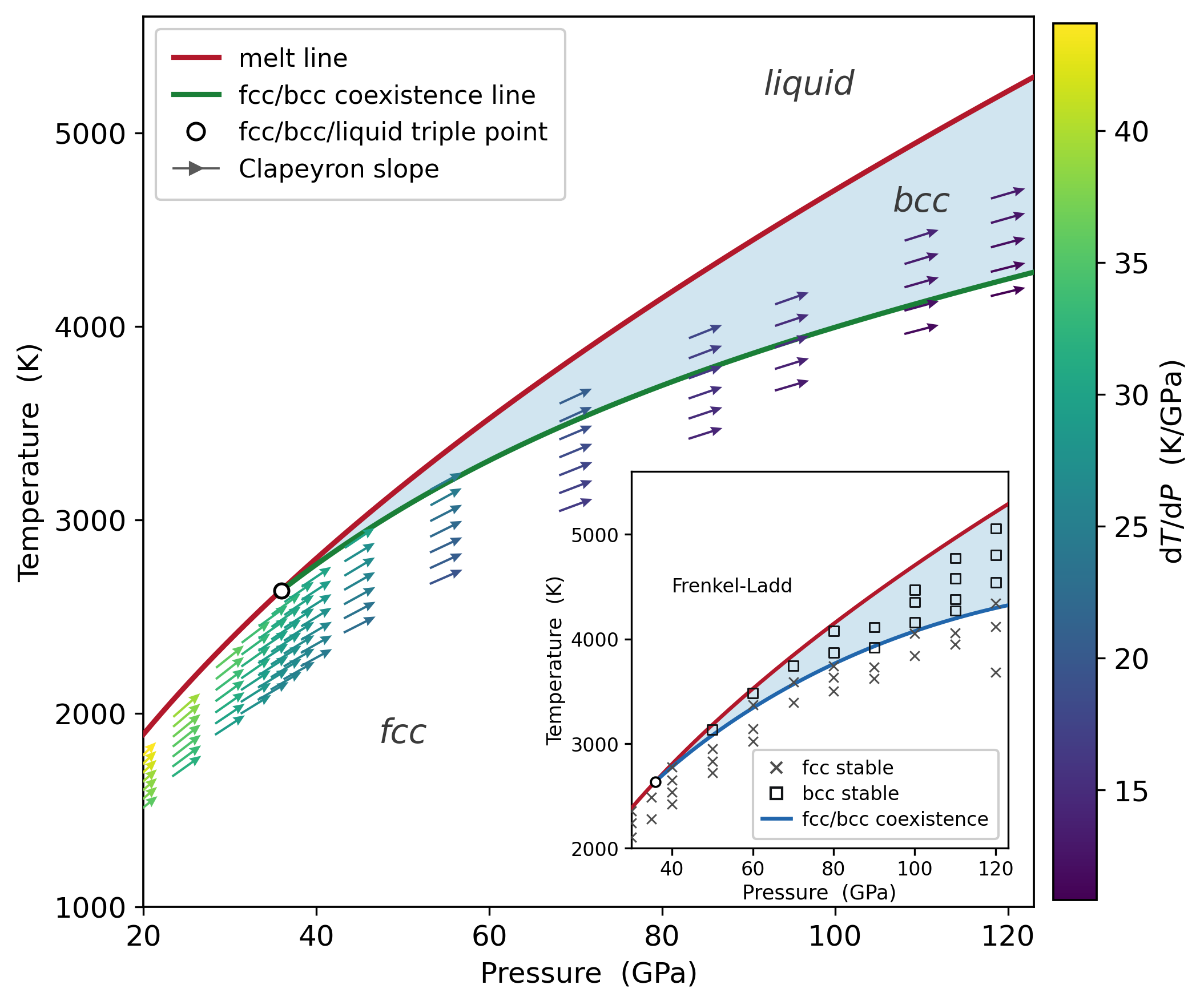}
\caption{{ 
The Clapeyron slope field of the 
fcc--bcc region of the phase diagram (Fig.~\ref{fig:phase}). Direction arrows, coloured by $\mathrm{d}T/\mathrm{d}P$) vary smoothly across the whole region.  The fcc/bcc coexistence line is the field line through the fcc/bcc/liquid triple point (circle) at the crossing of the fcc and bcc melting branches. Inset: independent Frenkel--Ladd free energies classify each $(P,T)$ state point as fcc- or bcc-stable, reproducing the same bcc field and coexistence line.}}
\label{fig:clapeyronfield}
\end{figure}

$\mathcal{C}(T,P)$ is a scalar field, but it is best represented as a unit vector field: the field line which passes through {\it any} point where $\Delta g=0$ defines the phase boundary (see Fig.~\ref{fig:clapeyronfield}). Thus, $\mathcal{C}(T,P)$ can be used to integrate along the phase line from any independently determined $\Delta g=0$ anchor.

\subsection{Gibbs--Helmholtz Integration}

The standard Gibbs--Helmholtz method gives free-energy changes for a single phase along an isobar.
The Gibbs--Helmholtz equation 
\begin{equation}
\left(\frac{\partial (g/T)}{\partial T}\right)_P = -\frac{h}{T^2}
\end{equation}
can be used to calculate the change in free energy of each phase along an isobar by numerical integration using NPT simulations over a range of $T$.
It is applied separately to each phase.
Therefore, at a fixed pressure $P$,
\begin{equation}
\frac{\Delta g(T,P)}{T}
=
\frac{\Delta g(T_0,P)}{T_0}
-
\int_{T_0}^{T}
\frac{\Delta h(T',P)}{T'^2}
\,dT'.
\label{eq:gh_integral_difference}
\end{equation}

Given $\Delta g$ between two phases at any point, one can locate the phase boundary where $\Delta g=0$ on the isobar. In practice, we did not use static relaxation as the anchor point because the integral diverges at $T=0$. One reliable anchor is a triple point on the melt line, where $\Delta g_{\alpha,\beta}(P,T)=0$. MD methods such as phase coexistence track these lines without calculating the absolute Gibbs free energy.
Free-energy differences between solid phases at finite $(T,P)$ can be calculated with methods such as Frenkel--Ladd\cite{frenkel1984new} or lattice-switch Monte Carlo\cite{bruce1997lattice,bruce2000lattice}, but they require carefully defined reference systems or bespoke switches and, in the case of an Einstein-crystal reference state, involve ``turning off'' quantum mechanics. In extremis, one can even use experimental data.

Using local Clausius--Clapeyron slopes to follow a coexistence line from a known anchor point is well established as Gibbs--Duhem integration\cite{KofkeMP1993,KofkeJCP1993}, with molecular dynamics applications to vapour--liquid and solid--liquid equilibria\cite{LisalVacek1996,LisalVacek1997}, and related single-simulation and alternative tracing schemes\cite{deKoning2001,Orkoulas2010}. Such integration is not self-correcting: errors in the anchor point or in the local slope accumulate along the line\cite{vantHof2006starting}. Plotting the full Clapeyron field addresses this concern: the smoothness of the field lines demonstrates the convergence of the underlying simulations, and the sensitivity of the traced boundary to the anchor can be read off directly from neighbouring field lines. For example, we found that for the fcc--hcp line at low $T$, the very small values of both $\Delta h$ and $\Delta v$ make the slope ill-conditioned and the method impractical.

\subsection{Anchoring points}

Both the Clapeyron field and Gibbs--Helmholtz integration determine only how the free-energy difference \emph{changes}. Each phase boundary must be anchored at one point where $\Delta g=0$ is known independently. The anchors used here are as follows. Two-phase coexistence simulations locate points on the melt lines directly. The fcc--bcc line is anchored at the bcc--liquid--fcc triple point, obtained from the crossing of the independently computed fcc and bcc melt lines. The fcc--hcp boundaries are anchored at $T=0$, where the Gibbs free energy reduces to the enthalpy of the statically relaxed lattice, supplemented by quasiharmonic zero-point contributions.

\subsection{Phase coexistence}

\begin{figure}    \centering
    \vspace{1em}
    \includegraphics[width=0.9\linewidth]{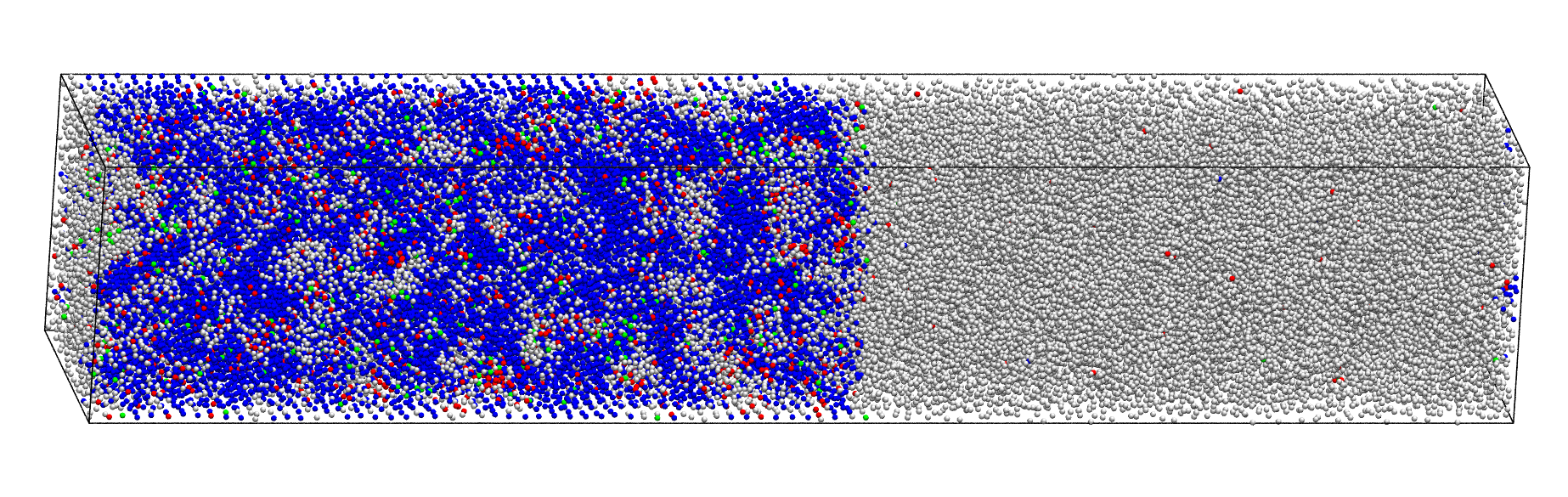}
    \captionsetup{width=0.9\linewidth}
    \caption{bcc--liquid phase-coexistence simulation in LAMMPS. Common-neighbour analysis identifies bcc atoms in blue; red and green atoms are classified as hcp and fcc, respectively.}
    \label{fig:pcbox}
\end{figure}

Melting lines for the fcc and bcc crystal structures were produced via solid--liquid phase-coexistence simulations using LAMMPS. The methodology involved constructing a simulation box with two shorter lateral dimensions and one longer dimension. The longer dimension was oriented perpendicular to the solid--liquid interface, as shown in Fig.~\ref{fig:pcbox}. This is necessary because periodic boundary conditions generate two such interfaces, so the box must be long enough in this direction to minimise interactions between them. The atoms were initialised in the crystal structure of interest across the full box.

\noindent

The system was first equilibrated around the estimated coexistence temperature and pressure in the isobaric--isothermal (NPT) ensemble for 200~ps. Next, two regions were defined, one for each half of the box volume. Dynamics for atoms assigned to one half of the box were paused while the other half was heated to approximately five times the equilibration temperature to produce a liquid. This was performed in the NPT ensemble over 200~ps. The melted region was then cooled back to the equilibration temperature over 400~ps in NPT. Finally, the full box, containing both solid and liquid regions, was run in NPH for 2~ns. The system was allowed to settle into an equilibrium coexistence state, with the solid--liquid interface free to move. The steady pressure and temperature of this final system were taken as the production data and averaged to form a point on the melting line. The fcc case used 123,120 atoms and the bcc case used 61,560 atoms.

\subsection{Quasiharmonic phonons}

At low temperatures, where the fcc--hcp free energy difference is far below the resolution of the methods above, we used lattice statics and quasiharmonic phonons. 

The fcc--hcp boundaries were calculated from static LAMMPS relaxation, including $c/a$, plus harmonic phonon free energies. The phonons were evaluated at the relaxed $T=0$ volume for each pressure with the ASE finite-displacement \texttt{Phonons} workflow and 0.01~\AA\ displacements in 864-atom (fcc) or 800-atom (hcp) supercells with sum-rule corrections\cite{ackland1997practical}. No imaginary modes were found. Explicit calculations spanned temperatures up to \SI{500}{K} and pressures up to \SI{140}{GPa}, with refined grids concentrated near the phase boundaries. For each phase and volume, phonon frequencies $\omega_i(V)$ were computed from the dynamical matrix of the two-body potential. The Helmholtz free energy is
\begin{equation}
F(V,T) = E_0(V) + \sum_i \left[ \frac{\hbar\omega_i(V)}{2} + k_B T \ln\left(1 - e^{-\hbar\omega_i(V)/k_B T}\right) \right],
\end{equation}
which includes the zero-point energy, and the Gibbs free energy follows as $G(P,T)=\min_V\left[F(V,T)+PV\right]$. Phase boundaries are located where $\Delta G_{\rm fcc-hcp}=0$. The quasiharmonic approximation is well suited to this problem: the fcc--hcp energy scale is set by distant-neighbour interactions and is tiny, while the anharmonic corrections neglected here are almost identical in the two stackings and largely cancel in the difference. 

\subsection{Free-energy verification}
As an independent check on the existence of the bcc phase, we computed absolute Gibbs free energies of the fcc and bcc solids by non-equilibrium Frenkel--Ladd integration\cite{frenkel1984new} as implemented in the calphy package\cite{menon2021automated,freitas2016nonequilibrium}. The Einstein crystal was used as the solid reference. The relative stability of the two solids is then read from the sign of $\Delta G_{\rm bcc-fcc}$.

\section{Results and Discussion}

\begin{figure}
    \centering
    \includegraphics[width=0.95\linewidth]{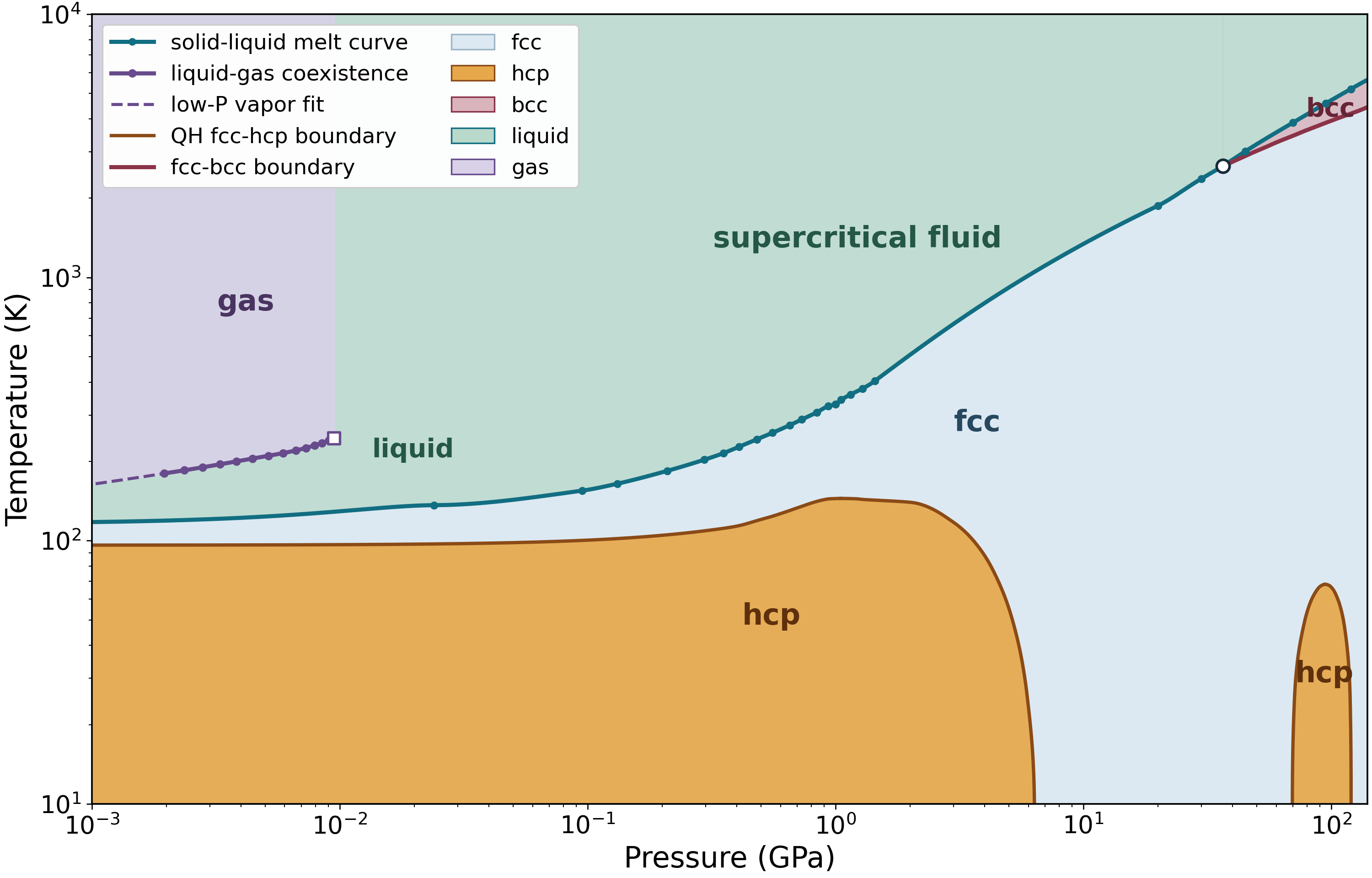}
\caption{
Phase diagram of the Tadah!Kr3b two-body potential. Melt lines are from two-phase coexistence, the fcc--bcc boundary from Gibbs--Helmholtz integration anchored at the bcc--liquid--fcc triple point, the liquid--gas line and critical point from NVT coexistence, and the low-temperature fcc--hcp boundaries from static relaxation and quasiharmonic free energies.}
\label{fig:phase}
\end{figure}

The calculated phase diagram is shown in Fig.~\ref{fig:phase}. It is assembled from four independent sets of calculations, presented in turn below. The fcc and bcc melting curves from two-phase coexistence cross at the bcc--liquid--fcc triple point, $(\SI{36}{GPa}, \SI{2630}{K})$. Since bcc does not extend to $T=0$, this is the only anchor of the fcc--bcc coexistence line, which is traced from it with the Clapeyron slope field and verified by Frenkel--Ladd free energies. The low-temperature fcc--hcp boundaries are anchored at $T=0$ by static relaxation and quasiharmonic free energies. The liquid--gas line and its critical point are determined directly from slab coexistence and require no anchor.
\begin{itemize}
\item fcc and bcc melting curves from two-phase coexistence,
\item anchoring points from melt curve coexistence and T=0 static relaxation,
\item Clapeyron field scans for fcc, bcc, and liquid, with Gibbs--Helmholtz integration for the fcc--bcc boundary,
\item direct coexistence liquid--gas slab calculations to find the boiling curve and critical point,
\item quasiharmonic free energies for the low-temperature fcc--hcp boundaries,
\item Frenkel--Ladd free energies verifying the existence of the bcc phase,
\item Ackland--Jones (AJ) structural diagnostics\cite{ackland2006applications}.
\end{itemize}






\subsection{Melt Curve}

\begin{figure}[htbp]
\centering
\includegraphics[width=0.92\textwidth]{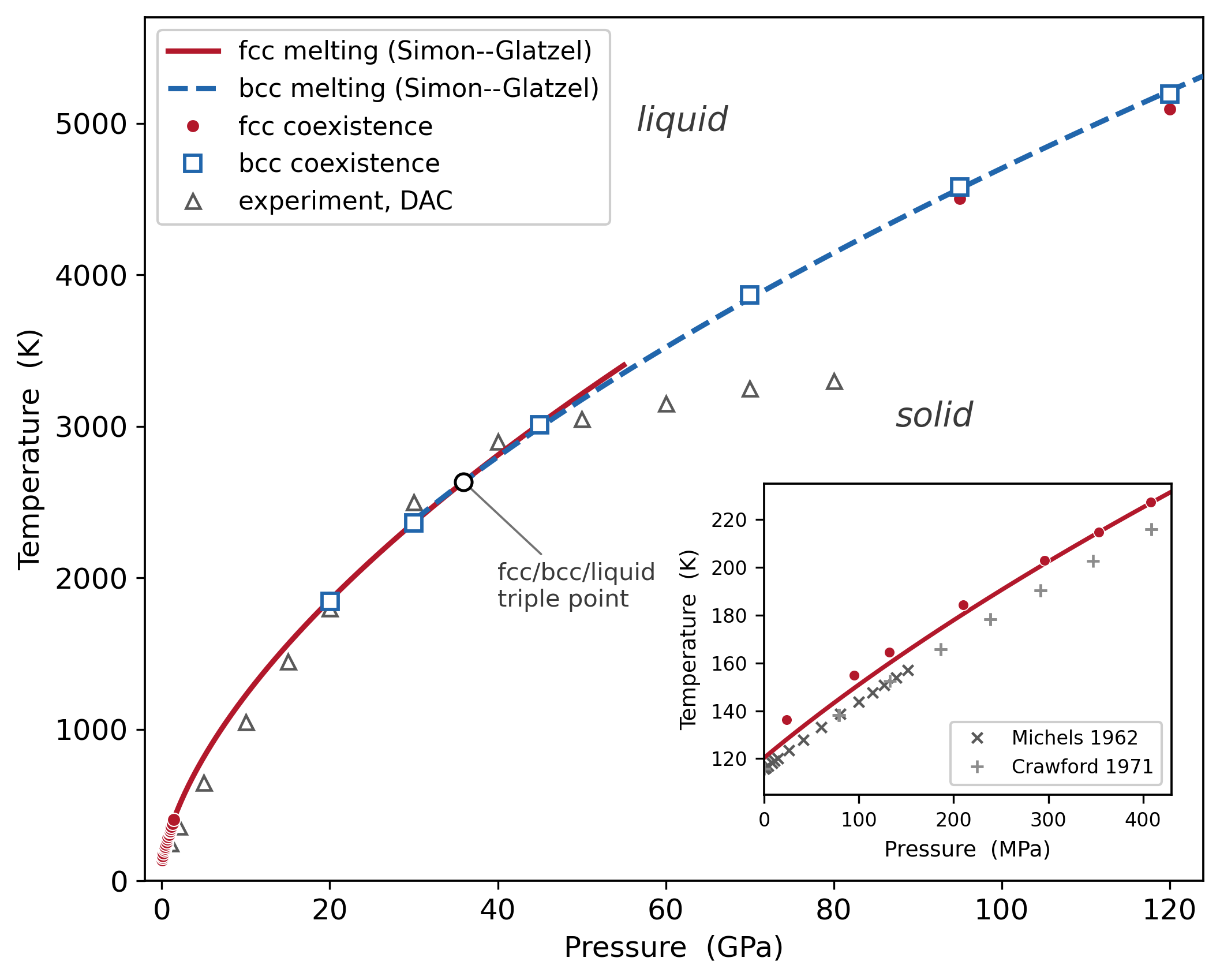}
\caption{{Melting curves of the fcc (circles) and bcc (squares) phases from two-phase coexistence, each fitted with a Simon--Glatzel form constrained to pass through the triple point as determined from the crossing of the two branches. Above the fcc/bcc/liquid triple point (open circle, $\approx\SI{36}{GPa}$, \SI{2630}{K}) the bcc branch melts at a higher temperature than fcc. Grey triangles are speckle-method diamond-anvil-cell measurements\cite{Boehler2001melting}. Inset: the low-pressure region compared with experimental melting data\cite{MichelsPrins1962,CrawfordDaniels1971}.}}
\label{fig:meltcross}
\end{figure}

The melt curve was tracked by a combination of two-phase coexistence calculations, which give points on the curves, and Clapeyron slope-field calculations, which give the slope at each point. Both fcc and bcc melt curves were computed across the full pressure range (Fig.~\ref{fig:meltcross}). The stable crystal structure is the one with the higher melting point, switching from fcc to bcc at a triple point near $(\SI{36}{GPa},\SI{2630}{K})$.

At low pressures the computed melt curve lies systematically above experiment\cite{MichelsPrins1962,CrawfordDaniels1971}, by about 13 K near the triple point (Fig.~\ref{fig:meltcross}, inset). This is a genuine shortcoming of the model, consistent with the tendency of two-body potentials fitted to dimer data to over-bind the close-packed solid. At high pressures the computed melt curve rises much more steeply than the laser-heated diamond-anvil-cell measurements\cite{Boehler2001melting}, which flatten above about 30 GPa. Some caution is needed in interpreting this discrepancy. Those experiments detected melting by the laser-speckle method, which is known to produce anomalously flat melting curves, and melting criteria based on x-ray detection of the liquid signal have yielded substantially higher melting temperatures in other systems\cite{Dewaele2010Ta}. A definitive test of the model at these conditions awaits x-ray melting data for krypton.

\subsection{fcc--bcc Line}

\begin{figure}[htbp]
\centering
\includegraphics[width=0.92\textwidth]{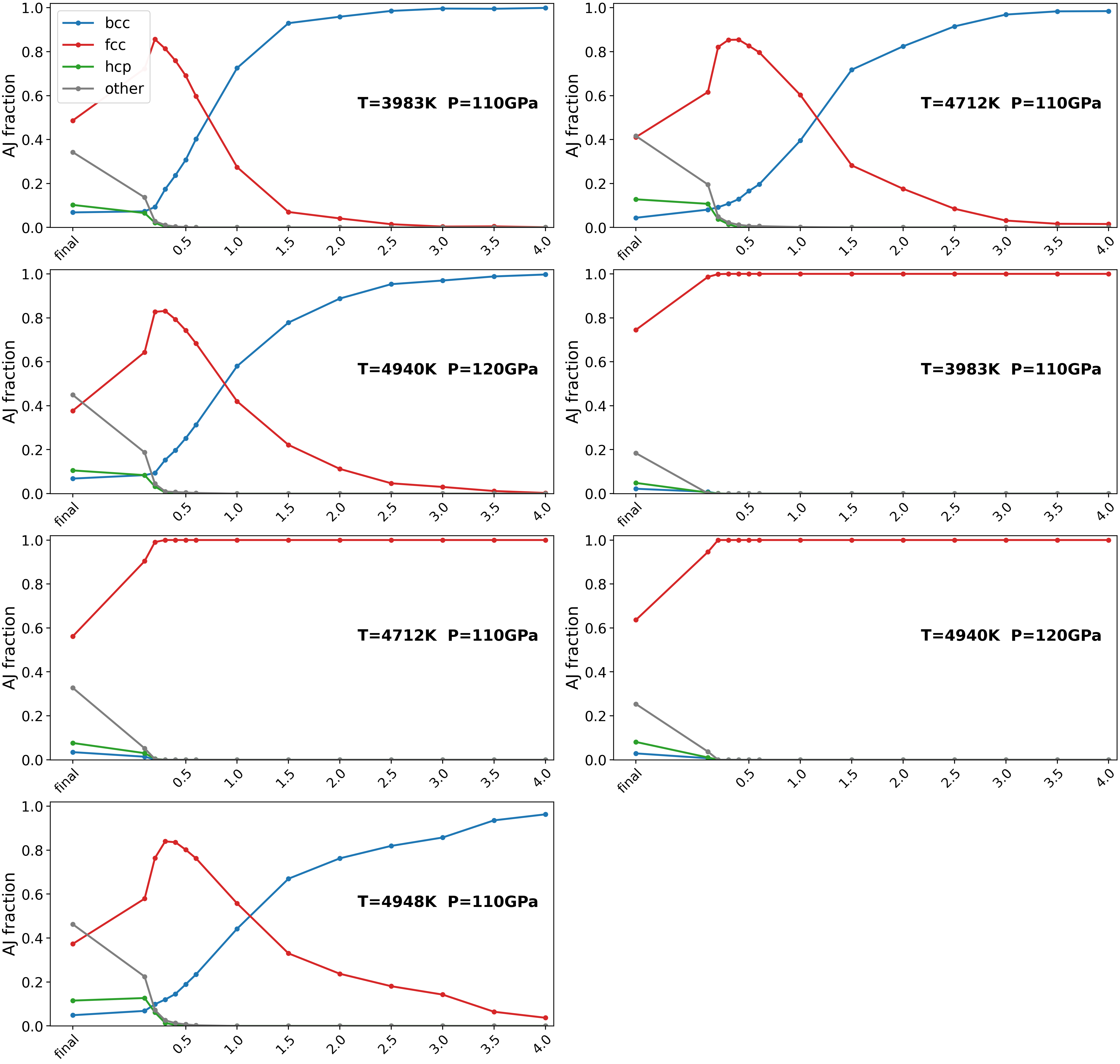}
\caption{AJ fractions\cite{ackland2006applications} for representative high-$T$, high-$P$ fcc- and bcc-tagged runs as a function of averaging window. Final configuration snapshots look mixed, with a preference for fcc in both fcc and bcc cells. However, the \SI{5}{ps}--\SI{10}{ps} averaged configurations clearly indicate the correct space- and time-averaged bcc phase.}
\label{fig:ajavg}
\end{figure}

The Clapeyron slope field and the Gibbs--Helmholtz integration are constructed from single-phase NPT simulations, so it is essential that metastable phases do not spontaneously transform. We used LAMMPS AJ analysis\cite{ackland2006applications} to determine phase stability.
The fcc structures remained well defined.
Instantaneous AJ classification based on snapshots from the high-$T$, high-$P$ bcc runs often looks strongly mixed among bcc, fcc, hcp, and ``other'' local environments. However, representative time-averaged structures show a different picture: averaging over a few ps gives predominantly bcc, and by \SI{10}{ps} the analysis is essentially pure bcc (Fig.~\ref{fig:ajavg}).

Time-averaged fcc configurations remain fcc, and the systematic behaviour of the Clapeyron slope field shows that the bcc and fcc simulations represent two separate phases.

The line starts at the triple point calculated from the intersection of the direct-coexistence melt curves.
We then calculated the enthalpies and volumes for the bcc-fcc pair using NPT simulations in the region around the phase line and converted these into the local Clapeyron slope field $\mathcal{C}(T,P)$ of eq.~\ref{eq:slope_field_definition}.

The fcc--bcc line in $(P,T)$ space was found by following the Clapeyron slope field.
To refine this at each pressure, we used the previous line temperature as a \emph{local} $\Delta G=0$ anchor and integrated the GH relation within the temperature stencil to reconstruct a locally anchored $\Delta G_{\mathrm{bcc-fcc}}(T,P)$ field.

The errors in integrating away from the triple point are cumulative, so
we used a Monte Carlo resampling of the refined GH block means to estimate the uncertainty. These errors start from zero (by definition) at the triple point, rising to about $\sigma=\pm 10$~K by 120~GPa.

The Frenkel--Ladd thermodynamic integrations provide an independent check on the Clapeyron slope-field construction, using the same potential but an entirely different methodology. At $T=0$ the enthalpy favours fcc/hcp at all pressures. Approaching the melt, the free-energy calculations independently find an entropically stabilised bcc band, with $\Delta G_{\rm bcc-fcc}$ changing sign within a few GPa of the triple point found by direct coexistence (Fig.~\ref{fig:clapeyronfield}, inset). Close to the melting point, the Einstein-crystal reference becomes ill-conditioned for bcc: soft springs allow atoms in the reference system to approach unphysically short separations, while stiff springs artificially constrain a phase which is already incipiently diffusive. Hcp free energies were not attempted because the fcc--hcp free-energy difference lies below the resolution of the Frenkel--Ladd method.

\subsection{fcc--hcp Line}

It is very unusual for a two-body potential to stabilise the same crystal structure across a full range of pressures, even at $T=0$. This is because the energy arises from shells of neighbours which sample different parts of $V(r)$ as the volume is reduced. Furthermore, in the present case, the energy difference between different stackings of close-packed structures depends only on third- and higher-neighbour interactions.

\begin{figure}
    \centering
    \includegraphics[width=0.95\linewidth]{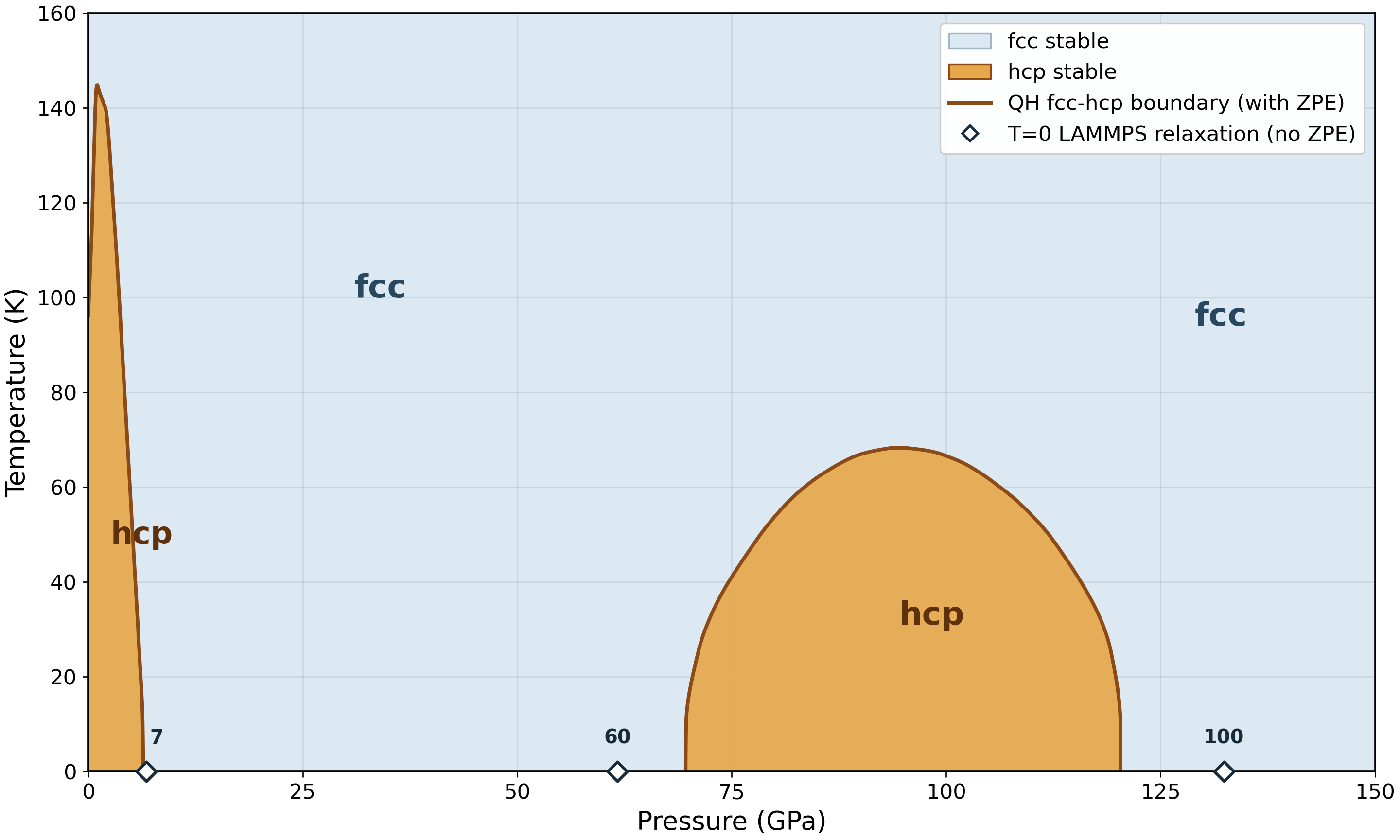}
    \caption{Low-temperature solid–solid phase diagram of the Tadah!Kr3b potential from
quasiharmonic free energies. Two hcp pockets, at low and at high pressure, exist and close with
increasing temperature. Diamonds mark the results from static relaxation, neglecting zero point energy. The melt curve is close to the y-axis and is not shown.
    \label{fig:qha}}
\end{figure}

We found that, on static relaxation, the present krypton potential undergoes a sequence of hcp--fcc--hcp--fcc transitions. The energy differences are extremely small, below \SI{0.1}{meV} per atom. This means that entropy plays an important role, with fcc having the higher entropy. We calculated the low-temperature phase diagram in the quasiharmonic approximation (Fig.~\ref{fig:qha}) and found that the fcc region expands its pressure range, superseding hcp at \SI{140}{K} in the low-pressure pocket and \SI{70}{K} in the high-pressure pocket, where zero-point energy also shifts the boundary significantly.

Experimental papers\cite{rosa2018effect,errandonea2002crystal} are sometimes misread as claiming an fcc--hcp phase transition under pressure. In fact, they show that fcc is the main phase throughout, with hcp as a minor component, and the data may be consistent with increasing amounts of stacking faults under pressure\cite{ackland2017quantum,pinsook2000atomistic}.
The "martensitic" pathway from fcc to hcp is generally attributed to stacking faults, so we take from experimental data only that the stacking-fault energy of solid Kr is low and that it deforms easily by the motion of partial dislocations, which generate hcp-like stacking-fault domains. Our simulations also show multiple stacking faults and regions closer to ABA (hcp stacking) than ABC (fcc).

Other potentials have similar sensitivities. The traditional model used for Kr, a Lennard--Jones potential, has at least 51 phase transformations as a function of the cutoff used at $T=0$, rising to several hundred when pressure is also considered. Once the cutoff exceeds $8\sigma$, hcp is established as the stable $T=0$, $P=0$ structure, consistent with our results here. Pressure and temperature effects favour fcc\cite{jackson2002lattice,wiebe2020phase}.

\subsection{Liquid--Gas Coexistence and Critical Parameters}
\begin{figure}[htbp]
\centering
\includegraphics[width=0.92\textwidth]{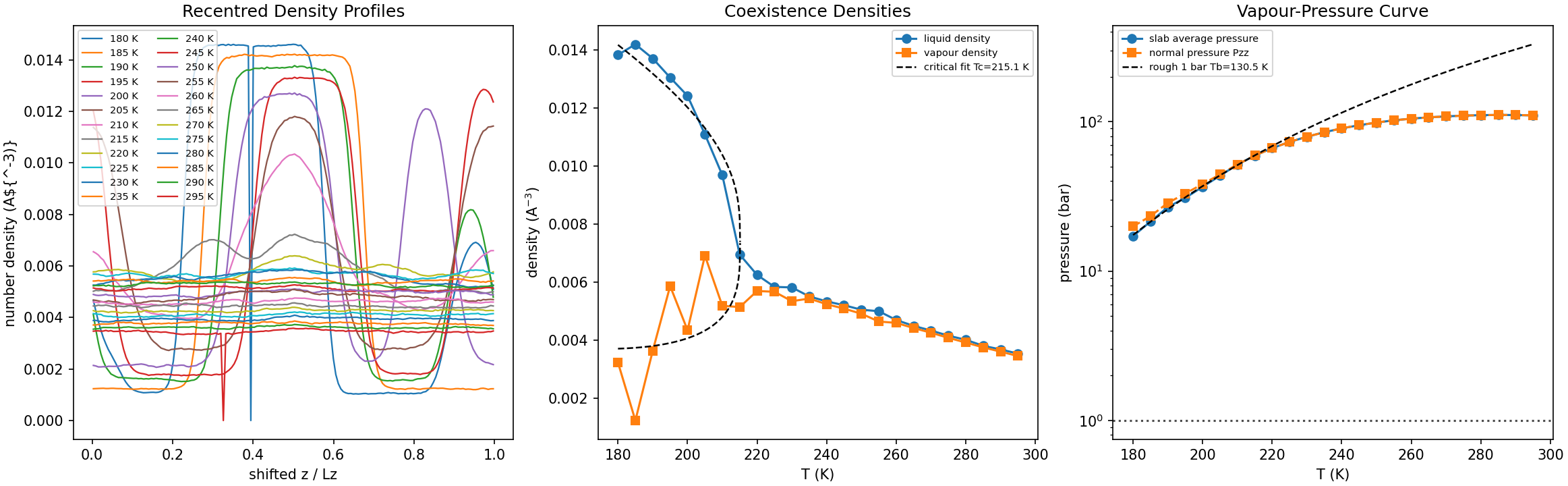}
\caption{Results from two-phase gas--liquid NVT calculations using LAMMPS with 11,232 atoms and a 1:1:4 aspect-ratio box. (a) Density profile along $z$, shifted to place the highest density at the centre, showing two distinct regions at low $T$ and becoming uniform above the critical point. (b) Densities extracted from the previous panel and plotted against $T$, with the coexistence curve giving $T_c=215.1$~K and $P_c=60$~bar. (c) Pressure extracted from the simulation, showing isotropy and a Clausius fit, $\ln P=a/T+b$, which extrapolates to an ambient-pressure boiling point of 130.5~K.
}
\label{fig:liqgas}
\end{figure}

The liquid--gas line was calculated with direct slab coexistence, using NVT molecular dynamics in elongated boxes containing a dense liquid slab plus vapour space. Below criticality, the liquid and vapour densities were extracted from recentred $z$-profiles by averaging plateau regions. The P$_{zz}$ component was used as the coexistence pressure. A summary of the results and process is shown in Fig.~\ref{fig:liqgas}.

\begin{figure}[htbp]
\centering
\includegraphics[width=0.4\textwidth]{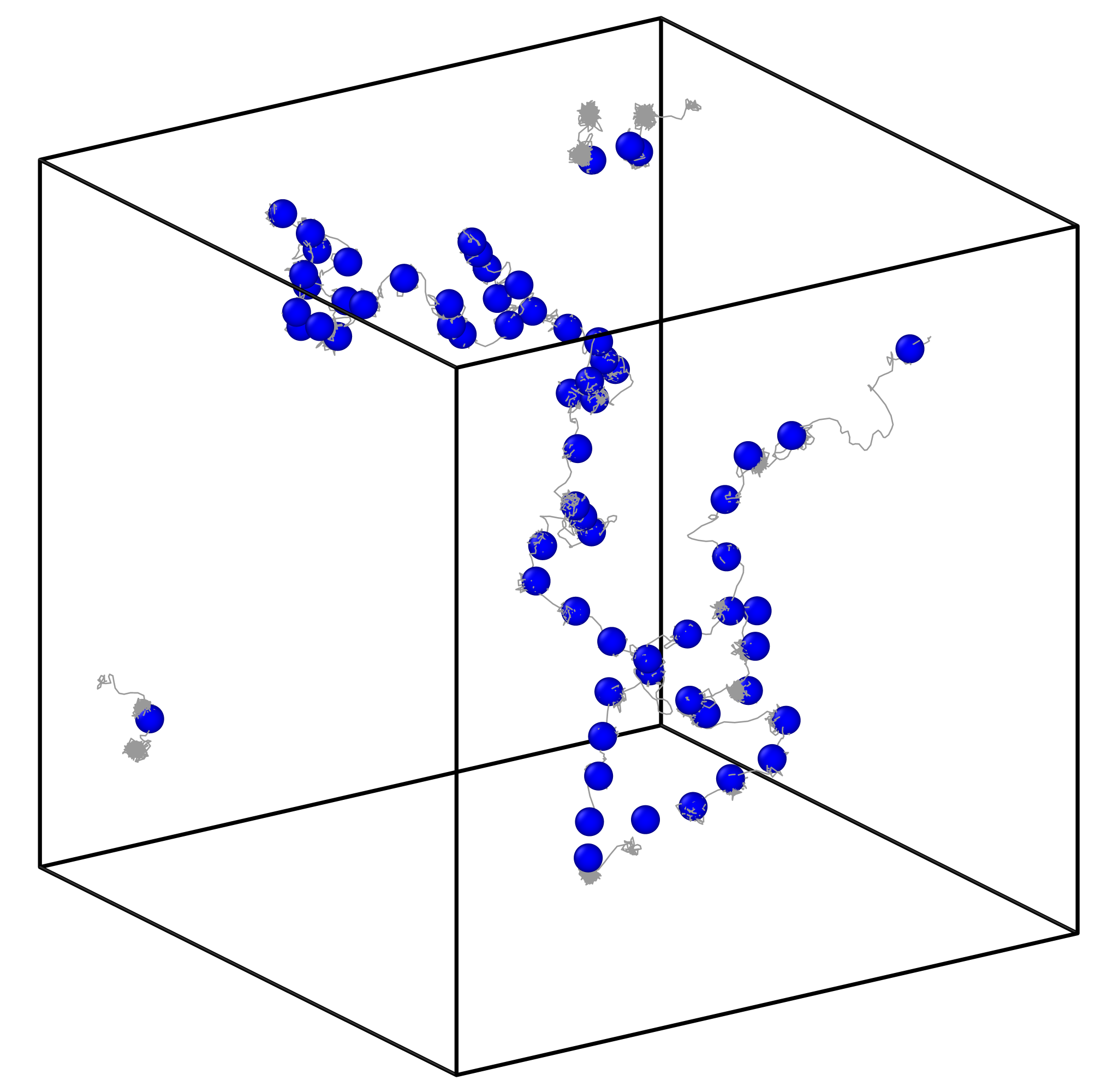}
\includegraphics[width=0.59\textwidth]{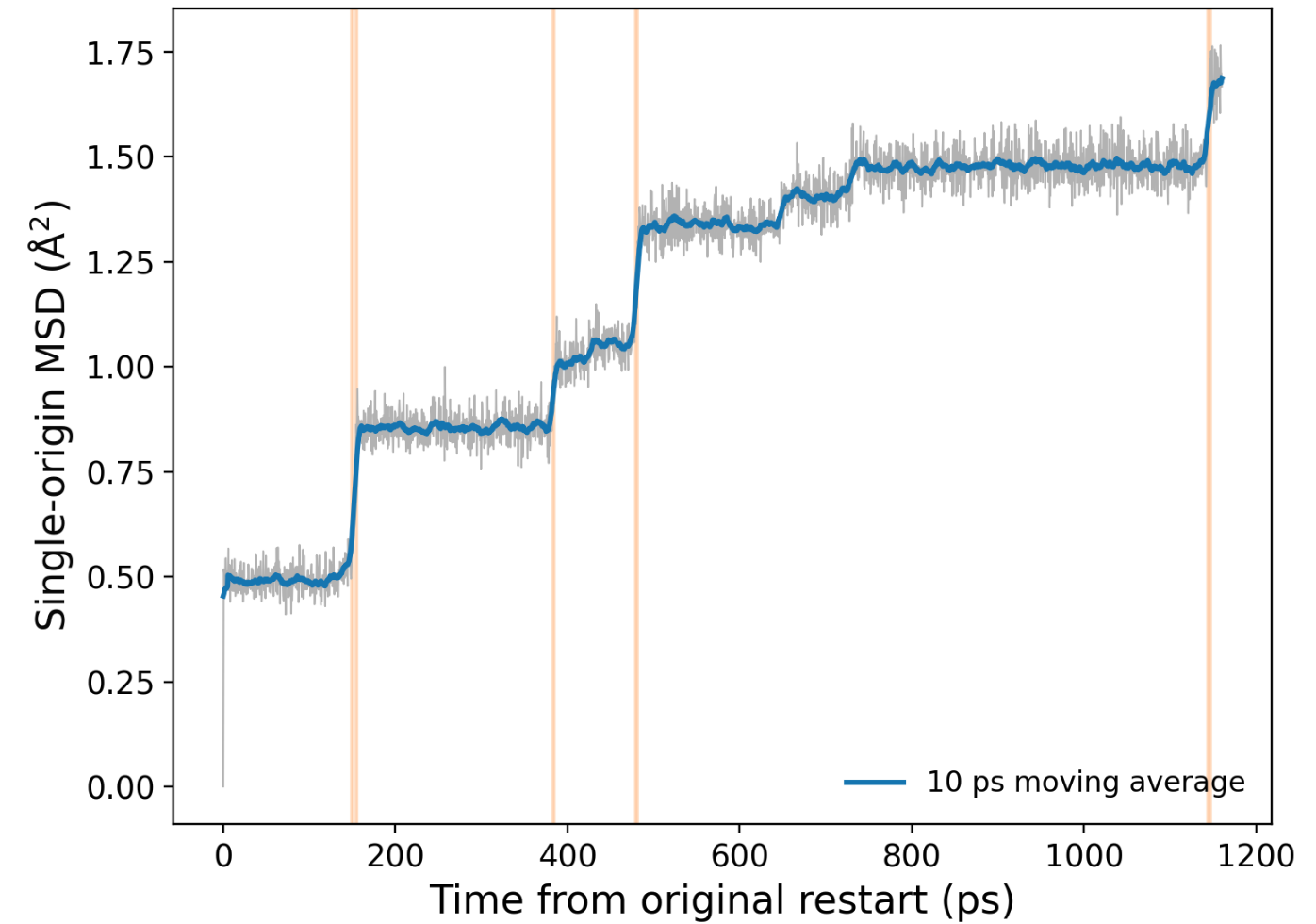}
\includegraphics[width=0.48\textwidth]{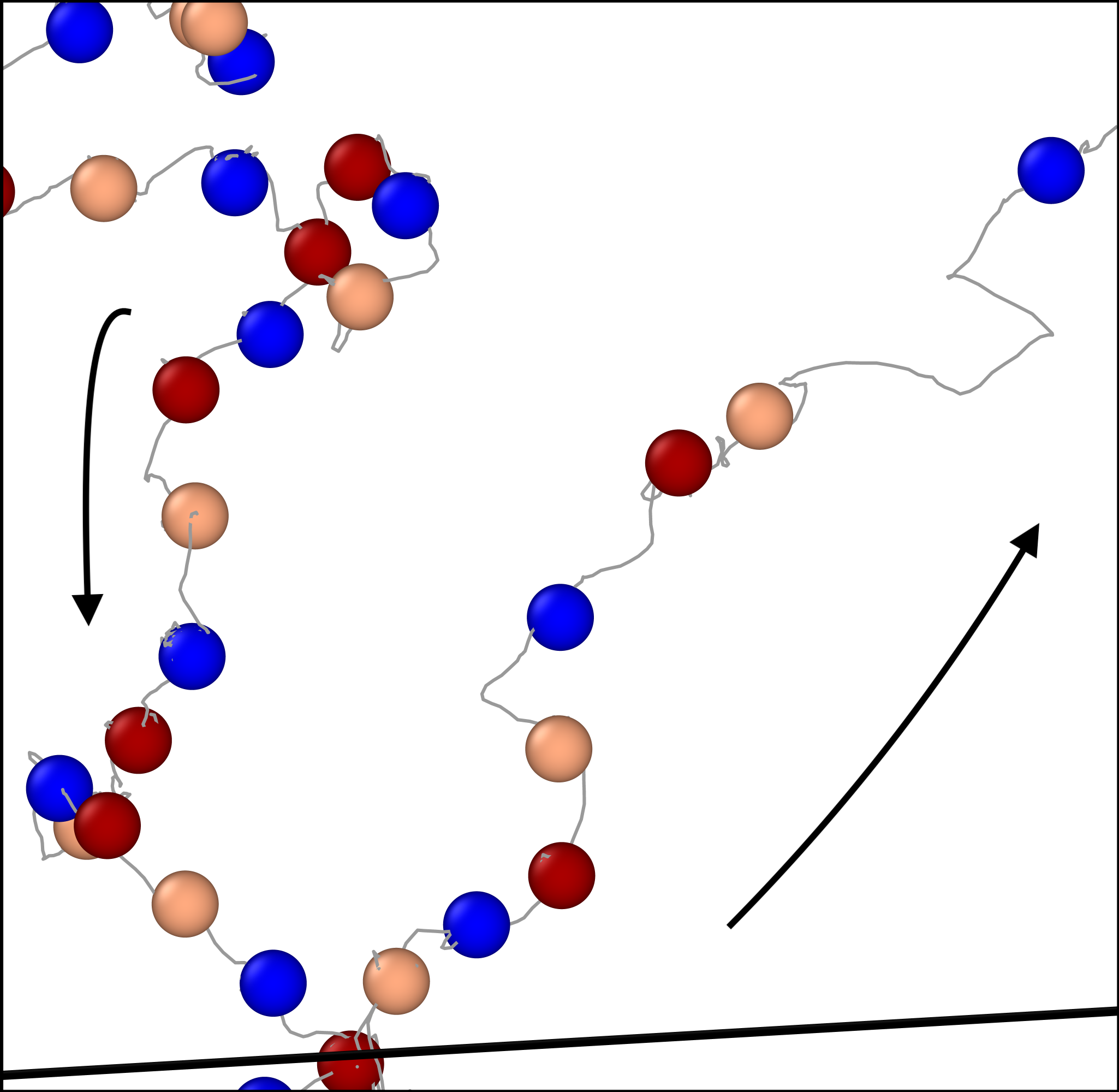}
\includegraphics[width=0.48\textwidth]{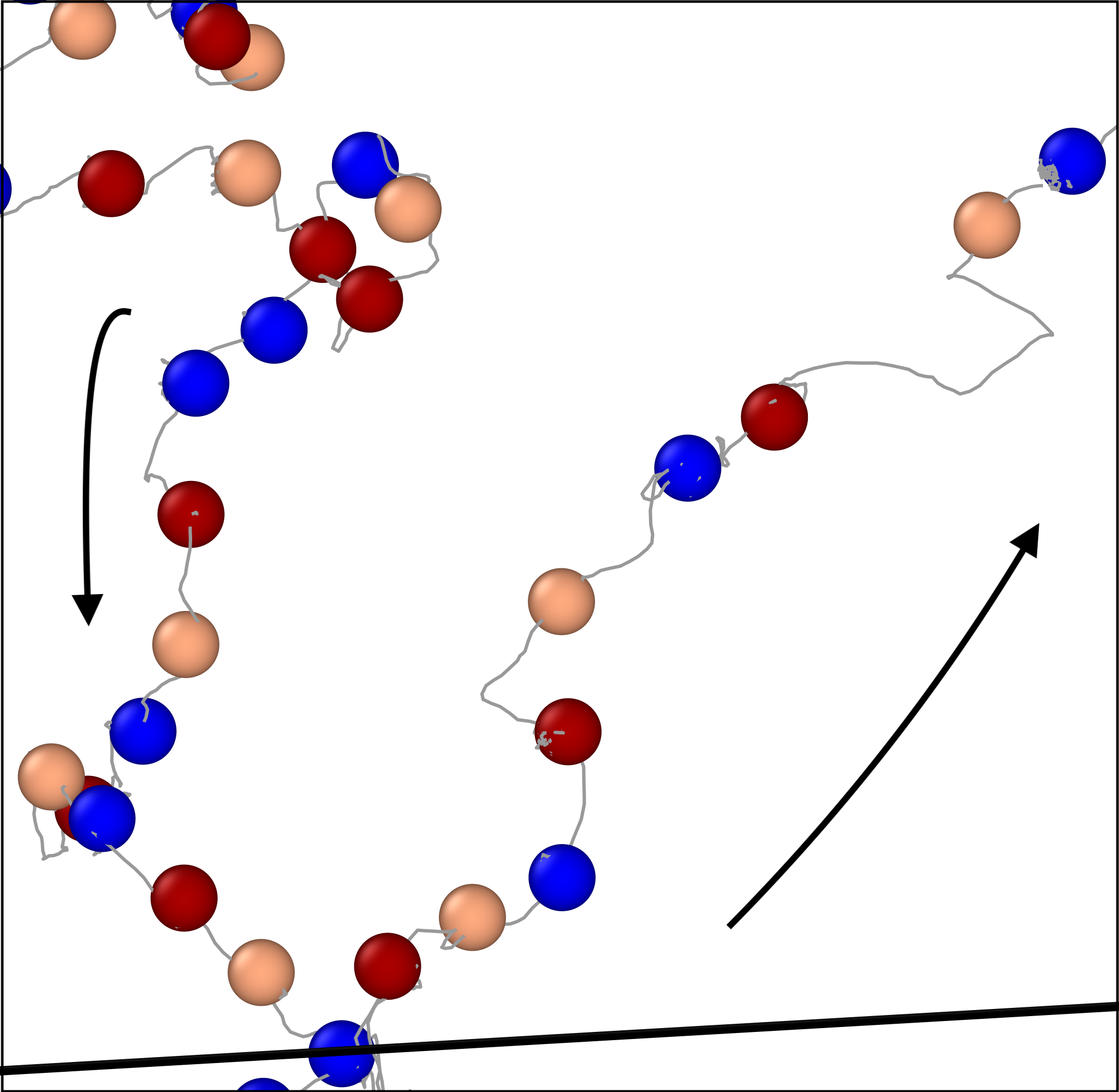}
\caption{Illustration of typical snake-like motion within bcc Kr. The thin grey lines show the trajectories followed by illustrated atoms  (top left) Trace of all the atoms which swapped lattice sites during a single greedy-snake event. (top right) Mean squared displacement from 1024 atoms bcc at 110GPa, 4948K.
(bottom left) randomly chosen red-blue colouring of a selection of the atoms in their initial positions (bottom right) Same colouring in the post-snake positions, showing that each atom has moved forward one site  (as defined by the arrows).
}
\label{fig:snakes}
\end{figure}

The critical-point estimate was obtained by fitting the distinct coexistence states to the order-parameter
\begin{equation}
\rho_l - \rho_v = A(T_c-T)^{\beta},\qquad 
\end{equation}
using the 3D Ising universality class value\cite{PelissettoVicari2002}
$\beta=0.326$ and fitting the diameter approximately as
\begin{equation}
\frac{\rho_l + \rho_v}{2} \approx \rho_c + B(T_c-T).
\end{equation}

The critical pressure was then estimated by fitting the coexistence pressure against temperature and evaluating that fit at the fitted $T_c$. The liquid--gas critical point obtained is
$T_c \approx \SI{215}{K}$,
$\rho_c \approx \SI{6.8e-3}{\per\angstrom\cubed}$, and
$P_c \approx \SI{60}{bar}$,
compared with the experimental values of \SI{209.48}{K} and \SI{55.25}{bar}\cite{LemmonSpan2006}.

To determine the statistical accuracy of the approach, we used bootstrap resampling over the distinct coexistence slab states, which gave approximate 16--84\% intervals of 215--220~K for $T_c$, $(6.4$--$7.5)\times10^{-3}$~\AA$^{-3}$ for $\rho_c$, and 59--67~bar for $P_c$.
These numbers reflect only fit sensitivity to the finite set of slab temperatures. They do \emph{not} include systematic errors from finite box size, interface broadening, or the pressure estimator itself.

The method becomes unstable at very low pressures: the diagonal components of the stress tensor become unequal because of fluctuations in a system with zero shear modulus, and the number of atoms in the gas region becomes small. We therefore fitted a Clausius curve, $\ln P=a/T+b$, to extrapolate from data between \SI{180}{K} and \SI{215}{K} back to ambient pressure ($a=-1374.8$, $b=10.48$). This extrapolation across an order of magnitude in pressure gives a remarkably good ambient-pressure boiling point of \SI{130}{K}, compared with the experimental value of \SI{119.7}{K}\cite{LemmonSpan2006} (Fig.~\ref{fig:liqgas}c). Interestingly, the Clausius curve departs from the measured $(P,T)$ data beyond the critical point.

\subsection{Greedy Snake Plasticity in bcc}

The stable bcc phase has higher enthalpy than fcc for example, at 110GPa/3983K we find a bcc enthalpy of 14.898eV/atom and fcc as 4.862eV/atom.  We also
find a mean squared displacement (MSD)  of 0.3557\AA$^2$ vs 0.2847\AA$^2$, indicating significantly higher entropy for bcc. 

In addition to the "vibrational" MSD, long runs in the bcc phase see occasional plastic events  which manifest as steps in the MSD (Figure \ref{fig:snakes}.  Detailed analysis shows these involve the correlated motion of a chain of atoms, preferentially along <111> directions. 
This mechaniism has been seen before
in simulations of bcc metals\cite{zepeda2004strongly,nguyen2006self,derlet2007multiscale} variously described as crowdion diffusion, chain motion and recently popularised as ``greedy  snakes"\cite{belonoshko2011ab,davis2011model}, which introduce a novel plasticity mechanism.

The Clausius Slope Field 
calculation  is smooth (Fig.\ref{fig:clapeyronfield}), and includes simulations where no such event occurred, so we conclude that the contribution of the snakes to the entropy is negligible.

\subsection{Comparison with MACE potential}

Pair potentials have intrinsic limitations which cannot be circumvented by fitting, but their simple form tends to protect against wildly unphysical behaviour.  On the other hand, more complex  machine-learned potentials  have the flexibility to better fit training data, but lack the physical grounding which protects against nonsense when training data is incomplete.  We illustrate the trade-off by comparing the Tadah! model to a published literature model: foundational MACE.  

\subsubsection{Clusters}

\begin{figure}[htbp]
\centering
\includegraphics[width=0.92\textwidth]{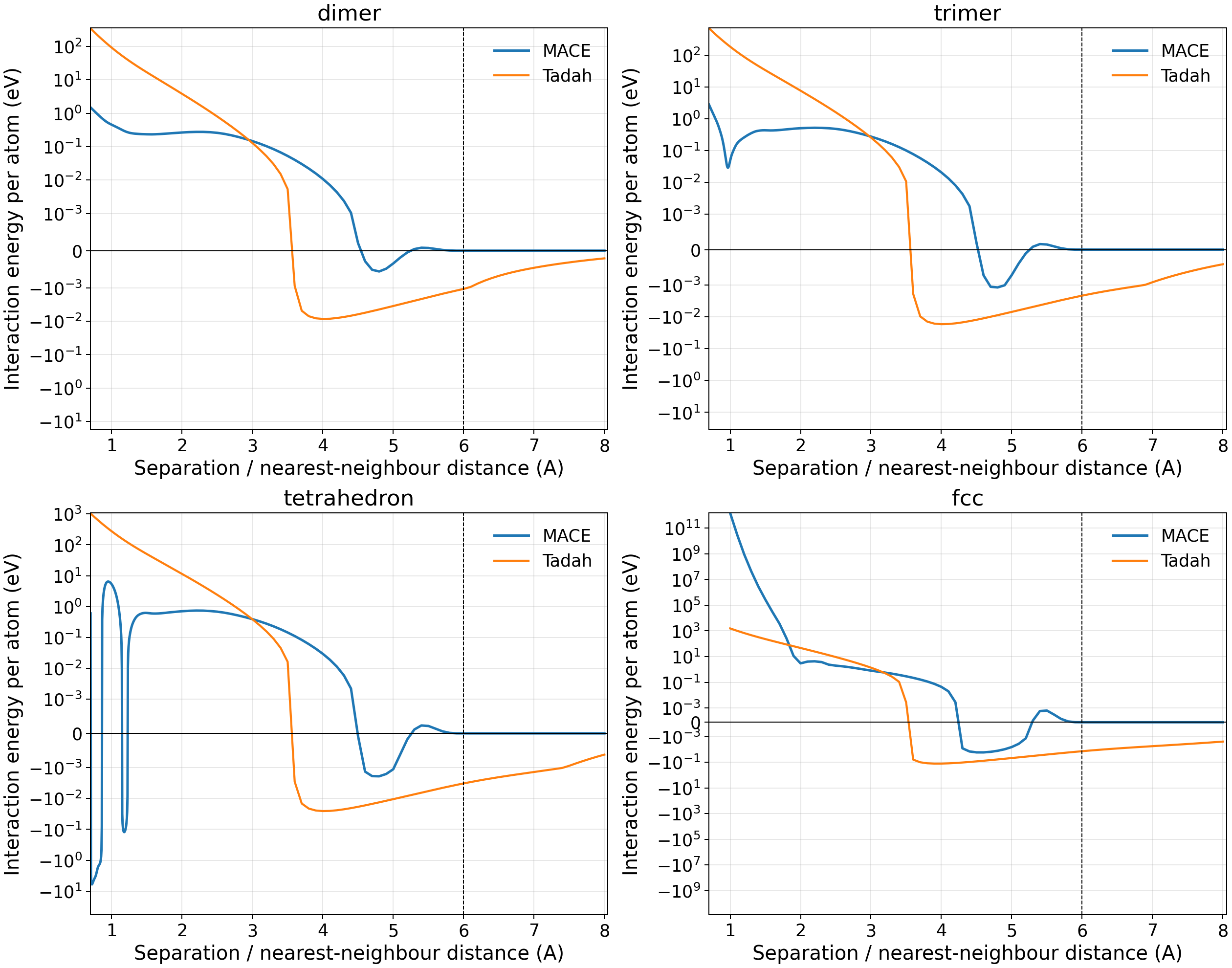}
\caption{Energies of various clusters on a symlog plot (linear below \SI{1}{meV}): a linear short range region indicates exponentially increasing repulsion.}
\label{fig:clusters}
\end{figure}

Clusters are the simplest structures for comparing two-body vs many-body effects. The dimer, equilateral triangle, and regular tetrahedron each have a unique bond length and only $60^o$ angles, so they sample just one point in the two-body potential. In Fig.~\ref{fig:clusters}, we compare the Tadah!Kr3b and MACE potentials.

The experimental dimer near-neighbour distance for Kr\cite{liu2020efficiency} is 4.034~\AA. The Tadah!Kr3b potential fits this well. This quantity has low weight in the MACE fit, so it is poorly described.
The trimer and tetrahedron in the Tadah! model are constrained by the functional form to be similar to the dimer, but this is consistent with the CCSD(T) training data: the trimer energy is very similar to three times the dimer energy.

MACE is trained on neither krypton nor explicit physical constraints, and so generates some unphysical behaviour. The tetrahedron has some very compact, highly bound states that are not suggested by the trimer and dimer data, indicating strong effects from the four-atom cluster. Indeed, these clusters are so stable that the most stable crystal structure is based on such motifs. By implication, these unphysicalities must be cancelled by similar unphysical behaviour in higher-order terms.

\subsubsection{Crystal structures}

The crystal structures at $T=0$ are a
good test for the usefulness of a potential, but there is no direct mapping from potential to the lowest energy structure.  Typically one proceeds by trial and error, testing plausible structures to find the lowest energy.

We saw above that the Tadah!Kr3b potential transforms between fcc and hcp with pressure; we investigate below the limitations of two-body potentials in avoiding such transitions. The interatomic separations change with pressure, which means that no individual feature of a two-body potential can stabilise a structure across a range of pressures: the whole function is important. Introducing angular terms helps, since the angles in a crystal structure are unchanged by compression. We might assume that MACE, informed about angles by the cluster expansion, might do better.

We ran structure search calculations on randomly generated cells: a number of codes already exist to do this, but we chose to generate our own trial cells at a range of pressures, ranking them by enthalpy and using spglib\cite{togo2024spglib} to characterise
the outputs. For Tadah!Kr3b, we were unable to find any stable structures other than fcc or hcp at any pressure. MACE shows metastable close-packed structures, favouring fcc over hcp.
However, for the most favourable structures, the MACE clusters exhibited a wide range of near-neighbour separations, 0.9--1.5~\AA.
Such structures are typically based on very stable clusters, as noted in the previous section.

\section{Conclusions}

We have computed the complete phase diagram of the Tadah!Kr3b two-body potential, fitted only to coupled-cluster dimer and trimer energies. Despite not being trained on condensed phase data, it reproduces the experimentally established features of krypton: the fcc solid, the melting and boiling curves, and a liquid--gas critical point close to experiment. It also makes two predictions with no experimental confirmation: an entropically stabilised bcc field below the melt curve, above the fcc/bcc/liquid triple point at $\approx\SI{36}{GPa}$, \SI{2630}{K}, and two pockets of hcp stability at low temperature. None of these features were included in the fitting.

The bcc phase  does not occur for Lennard--Jones, and we can trace the cause to the Tadah!Kr3b potential having a softer short-range repulsion than Lennard--Jones, which also give a better fit to equation of state data\cite{iwasaki2025accurate}.
The idea of a bcc region close to the melt curve is widespread, beginning with the famous  paper ``Should all crystals be bcc?" by Alexander and McTague\cite{alexander1978should}. While this title falls foul of Betteridge's Law, the continuation of the fcc-bcc line metastably into the melt region  suggests the possibility that
in comparison of crystal phases only, bcc may be the most stable in all cases at high temperature.

Our calculated melt curve tracks the experiment \cite{Boehler2001melting} well up to \SI{50}{GPa}.  Beyond that, it does not reproduce the reported anomaly which occurs close to where we observe a bcc phase. The experimental diagnostic is the change in the ``speckle pattern" arising from changes in  surface roughness and refractive index. Our simulations suggest that the greedy-snake defects increase plasticity and enable rapid changes in surface morphology which could be mistaken for melting, and that this experimental speckle-disappearance line should be interpreted as the fcc--bcc transition\cite{belonoshko2006xenon}.

 Although there are experimental claims of hcp Kr coexisting with fcc Kr\cite{errandonea2002crystal,shimizu2009high}, the data leading to those conclusions can equally well be attributed to stacking faults\cite{ackland2017quantum}. We suspect that the hcp-Kr pockets are highly sensitive to details of the fitting. 

The closing of the hcp pockets shows that fcc has higher entropy than hcp, and this, rather than the fcc--hcp enthalpy difference, dominates the phase diagram. This is also true for model systems such as Lennard--Jones and hard spheres, and follows from the quasiharmonic approximation and Landau-theory analysis\cite{alexander1978should}.

Overall, the study shows that the analysis and simulation methods 
using machine-learning methods for sophisticated nonlinear optimisation of fast, simple potentials can reveal rich details about fundamental condensed matter systems.  It emphasises the crucial role of physical form and fitting above mathematical flexibility, and the power of machine learning to optimise non-linear parameters.
It also demonstrates the need for multiple simulational techniques to cover different phase boundaries.
It sheds new light on the high pressure melting of krypton via a mechanism which may provide general resolution of the discrepancy between diffraction and speckle detection of the melt line. 

\section*{Supporting Information}

All calculations in this paper were done with ASE-based  python scripts, calling LAMMPS and  which are available on reasonable request.
For the purpose of open access, the author has applied a Creative Commons Attribution (CC BY) licence to any Author Accepted Manuscript version arising from this submission.
\section*{Acknowledgments}
MK acknowledges funding from an EPSRC Doctoral Prize Fellowship.  the authors have no conflict of interest.

\bibliography{GH}

\section*{TOC Image}
\includegraphics[width=0.92\textwidth]{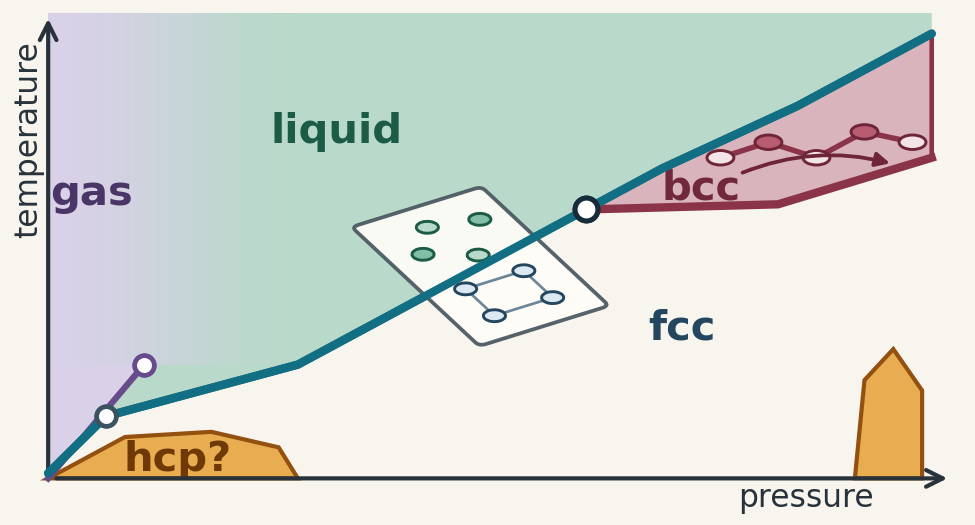}

\end{document}